\documentclass[12pt]{article}
\usepackage{a4wide,amsmath,amssymb,graphicx}
\usepackage{colordvi,gnuplotcolors}

\newcommand{\gev}{\text{GeV}}
\newcommand{\tev}{\text{TeV}}
\newcommand{\fb}{\text{fb}}

\newcommand{\TE}[1]{\cdot 10^{#1}}
\newcommand{\E}[1]{10^{#1}}

\newcommand{\MM}{\mathcal{M}}
\newcommand{\sqrts}{\sqrt{s}}
\newcommand{\BR}{\text{BR}}

\newcommand\Sq{\tilde q}
\newcommand\Stop{\tilde t}
\newcommand\Sbot{\tilde b}

\newcommand{\mhpm}{m_{H^\pm}}

\newcommand{\MSQ}{M_{\tilde Q}}
\newcommand{\MSU}{M_{\tilde U}}
\newcommand{\MSD}{M_{\tilde D}}
\newcommand{\MSL}{M_{\tilde L}}
\newcommand{\MSE}{M_{\tilde E}}
\newcommand{\mh}{m_{h^0}}

\newcommand{\mA}{m_{A^0}}
\newcommand{\MSf}{M_{\tilde f}}

\newcommand{\ee}{e^+ e^-}

\newcommand{\WH}{W^{\pm} H^{\mp}}
\newcommand{\hmp}{H^{\mp}}
\newcommand{\wpm}{W^{\pm}}
\newcommand{\tb}{\tan\beta}

\newcommand\eg{e.g.\ }
\newcommand\ie{i.e.\ }
\newcommand\rd{\mathrm{d}}
\newcommand\ri{\mathrm{i}}
\renewcommand\Re{\mathop{\mathrm{Re}}}

\begin{document}

\begin{titlepage}
\begin{flushright}
{IPPP/06/67\\
DCPT/06/134\\
MPP-2006-128\\
hep-ph/0610079}
\end{flushright}
\vfill
\begin{center}
{\bf\large On MSSM charged Higgs-boson production in association with\\[.2cm]
	an electroweak $W$ boson
	at electron--positron colliders}\\[2cm]
{\bf\large Oliver Brein
}\\[0.2cm]
{\normalsize Institute for Particle Physics Phenomenology,\\
University of Durham, DH1 3LE, Durham, United Kingdom}\\[.8cm]
{\bf\large Thomas Hahn
}\\[0.2cm]
{\normalsize Max-Planck-Institut f\"ur Physik\\
D-80805 M\"unchen, Germany}
\end{center}
\vfill
\begin{abstract}
We present a calculation of the cross section for the process 
$\ee\to\WH$ in the Minimal Supersymmetric Standard Model (MSSM)
and the Two Higgs Doublet Model (THDM).
We study the basic features of the MSSM prediction for some
distinctive parameter scenarios. We find large effects
from virtual squarks for scenarios with large mixing in the stop sector
which can lead to a cross section vastly different from a 
THDM with identical Higgs-sector parameters.
We investigate this interesting behaviour in more detail 
by thoroughly scanning the MSSM parameter space for regions of 
large cross section. 
For a charged Higgs boson too heavy to be pair-produced 
at such a machine, it turns out that a large MSSM cross section with 
a good chance of observation is linked to a squark mass scale below $600\,\gev$ 
and a considerable amount of mixing in either the stop and sbottom sector.
\end{abstract}
\vfill
\end{titlepage}

\section{Introduction}

The discovery potential of the CERN Large Hadron Collider (LHC) 
should be sufficient to resolve the issue of the existence 
of a neutral Higgs boson \cite{atlas-cms-summary}.  
An electron--positron collider would, however, serve as an ideal tool to
measure the properties of neutral Higgs bosons very precisely \cite{LC}.
To discover a charged
Higgs boson, $H^\pm$, on the other hand, is much harder at the LHC
\cite{atlas-cms-summary,Hpm-discovery-reach}, especially if the $H^\pm$
is substantially heavier than the top quark.  The reason for this is the
dominant decay of the $H^\pm$ bosons into heavy quarks ($t\bar b$ or
$\bar t b$) which leads to hadronic signatures that are hard to
distinguish from QCD background events and which forces one either to
rely on the less probable decay $H^\pm \to \tau\nu$ or to accumulate a
lot of statistics during several years of running at high luminosity in
order to claim discovery.  Actually, the strong dependence of the
production cross sections on the ratio of vacuum expectation values in
the Higgs sector, $v_2/v_1 = \tan\beta$, makes it almost impossible to
discover the charged Higgs boson at the LHC in a wedge-shaped region of
intermediate $\tan\beta$-values in the $\tb$--$\mhpm$ plane if it is
heavier than the top-quark \cite{atlas-cms-summary,Hpm-discovery-reach}. 
This non-discovery range includes $4\lesssim \tb \lesssim 12$ at $\mhpm
= 250\,\gev$ and widens continuously to \eg $2\lesssim \tb \lesssim 40$
at $\mhpm = 650\,\gev$ \cite{Hpm-discovery-reach}.

At an $\ee$ collider with a centre-of-mass energy $\sqrt s$, the main
production process for charged Higgs bosons is pair production ($\ee \to
H^+ H^-$) which is mediated mainly via photon- and $Z$-exchange in the
s-channel.  The pair-production cross section is almost independent of
$\tan\beta$ and, owing to the colourless initial state, Higgs-boson
signatures from decays into heavy quarks are not obscured by enormous
amounts of QCD background events, like at the LHC.  Therefore, detection
of charged Higgs bosons via pair production should be possible in the
whole $\mhpm$--$\tb$ plane up to a mass limit somewhat below $\sqrt
s/2$.  If the collider energy is not sufficient for pair production (\ie
$\sqrt s < 2 \mhpm$), the only way to extend the search for the charged
Higgs boson to higher masses is to try to observe its single production
in association with lighter particles.  The most relevant processes
which have been studied in the literature are the tree-level processes
$\ee\to H^\pm\tau\nu$ \cite{H-tau-nu,single-H-overview} and $\ee\to
H^\pm t b$ \cite{H-t-b,single-H-overview}, and the loop-induced
processes $\ee\to H^\pm e \nu$
\cite{H-e-nu,H-e-nu-SUSY,single-H-overview} and $\ee\to W^\pm H^\mp$
\cite{logan-su1,zhu-eewh,shinya-eewh,arhrib,obr-susy02,single-H-overview}. 
Most of the studies are done in the framework of a Two Higgs Doublet
Model (THDM) with parameters fixed at MSSM values, except for
Refs.~\cite{shinya-eewh,arhrib} which study $\WH$ production in the
general THDM, Refs.~\cite{logan-su1,obr-susy02} which study the same
process in the full MSSM, and Ref.~\cite{H-e-nu-SUSY} which includes
virtual sfermion contributions for the cross section prediction of the
$H^\pm e \nu_e$ final state.

If the collider energy does not allow for pair production and $\sqrt s
> m_W + \mhpm$, the $W^\pm H^\mp$ production process can become the
dominant source of charged Higgs bosons, depending on $\tb$.  
In this case, as the process is loop-induced, the signal rate is much smaller
than the typical pair-production rate.
Even if $H^+H^-$ production is
the main production process at an $\ee$ collider, it would still be very
rewarding to measure the $W^\pm H^\mp$ production cross section as well. 
While the $H^+H^-$ cross section is determined at tree-level by the
gauge couplings to photon and $Z$-boson, the $W^\pm H^\mp$ cross section
is loop-induced and thus depends already at leading order on the virtual
particles in the loops.  Therefore, with a cross section measurement of
the latter process one would have access to information about the
underlying model.

This paper presents a calculation of the MSSM prediction for the process
$\ee\to W^+ H^-$.  By the time this work was finalized, another
calculation of the same prediction appeared \cite{logan-su1}. We carried
out a detailed comparison \cite{su-private} of our results with those of
Ref.~\cite{logan-su1} which lead to agreement\footnote{
	As a result, the formulae in \cite{logan-su1} are now 
	checked by an independent calculation.  According to 
	\cite{su-private} the results agree, if in Eq.~(C14)
	of Ref.~\cite{logan-su1} the tensor
	coefficient $D_{23}$ in the coefficient of ${\cal A}_6 g_R^W 
	g^L_H$ is replaced by $2 D_{23}$.}.
In the framework of the Two Higgs Doublet Model (THDM), the 
cross section prediction for $\ee \to W^+ H^-$ is well-known
\cite{zhu-eewh,shinya-eewh,arhrib} and we could reproduce the results of
Ref.~\cite{arhrib} in particular.

In Section \ref{sect:theprocess} we present the process $\ee\to W^+ H^-$
with the contributing Feynman diagrams in the MSSM and 
some details about our calculation.
Section \ref{sect:results} presents our numerical results 
in two stages. 
In stage one, we examine the basic features of the $\WH$ cross section
for a $500\,\gev$ and $1\,\tev$ $e^+ e^-$ collider for two distinct
MSSM parameter scenarios and compare the results also to predictions 
of a THDM with the same Higgs sector parameters.
This enables us to exemplify the range of influence of virtual superpartners
on the cross section. The effect of polarized electron and positron
beams on the cross section is discussed briefly followed by a demonstration
of the decoupling of superpartners with increasing supersymmetry-breaking scale.
In stage two, we present results of a thorough
scan of the MSSM parameter space for regions
of large, possibly observable, cross section and try to understand
for which parameter scenarios they arise.
Section \ref{sect:conclusion} contains our conclusions.
Some results of this calculation were already reported in \cite{obr-susy02}.

\section{$\ee \to W^+ H^-$ in the MSSM}
\label{sect:theprocess}

\subsection{Kinematics}
\label{sect:kinematics}

We study the reaction
\begin{equation*}
  e^+(\bar k,\bar\sigma) + e^-(k,\sigma) \to
  W^+(p,\lambda) + H^-(\bar p)\,,
\end{equation*} 
where $k$ and $\bar k$ denote the momenta of the initial-state electron
and positron, $p$ and $\bar p$ the momenta of the final-state gauge
boson $W^+$ and Higgs boson $H^-$.  Additionally, the electron, 
positron, and $W$-boson are characterized by their spin polarization 
$\sigma, \bar\sigma (=\pm\frac 12)$ and $\lambda (=0,\pm 1)$.  
Neglecting the electron mass, the kinematical invariants $s = (k + \bar 
k)^2$, $t = (k - p)^2$ and $u = (k - \bar p)^2$ fulfil the relation
\begin{equation}
  s + t + u =  m_W^2 + \mhpm^2\,.
\end{equation}
We assume unpolarized electron and positron beams.  Thus, the
differential cross section summed over spin polarizations of the 
final-state $W$-boson reads
\begin{equation}
  \frac{\rd\sigma}{\rd t} = \frac 1{16\pi s^2} \:
    \sum_{\lambda = 0,\pm 1}
    \frac 14\sum_{\sigma,\bar{\sigma} = \pm\frac 12}
    \bigl|\MM_{\sigma\bar\sigma\lambda}\bigr|^2
\end{equation}
with the helicity amplitudes $\MM_{\sigma\bar\sigma\lambda}$
of the process.  The total cross section is evaluated by numerical 
integration over the kinematically allowed $t$-range:
\begin{equation}
  \sigma_{\ee \to W^+ H^-}(s) =
    \int_{ t_{\text{min}}(s) }^{ t_{\text{max}}(s) }
    \rd t\,\frac{\rd\sigma}{\rd t} (s,t)\,.
\end{equation}

\subsection{Feynman graphs}
\label{feynmandiags}

In the MSSM and THDM all leptons couple gauge-invariantly to one of the
two Higgs doublets and therefore also to the physical Higgs bosons. 
Thus, there is a non-vanishing tree-level amplitude for the process
under study.  The Feynman graphs on tree-level consist of three graphs
with $s$-channel exchange of neutral Higgs bosons ($h^0$, $H^0$, $A^0$)
and one with a $t$-channel exchange of a neutrino.  The Yukawa couplings
which appear in all tree-level graphs are $\propto m_e/m_W \approx
6\TE{-6}$ and suppress the tree-level contribution strongly.  The
tree-level amplitude can thus safely be neglected and consequently there
are no photon bremsstrahlung corrections at leading order.

\begin{figure}
\begin{center}
\centering{\includegraphics{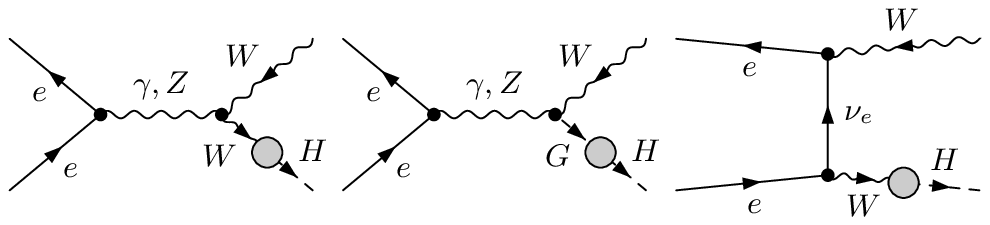}}

(a) graphs with self-energy insertions\\[.5cm]

\centering{\includegraphics{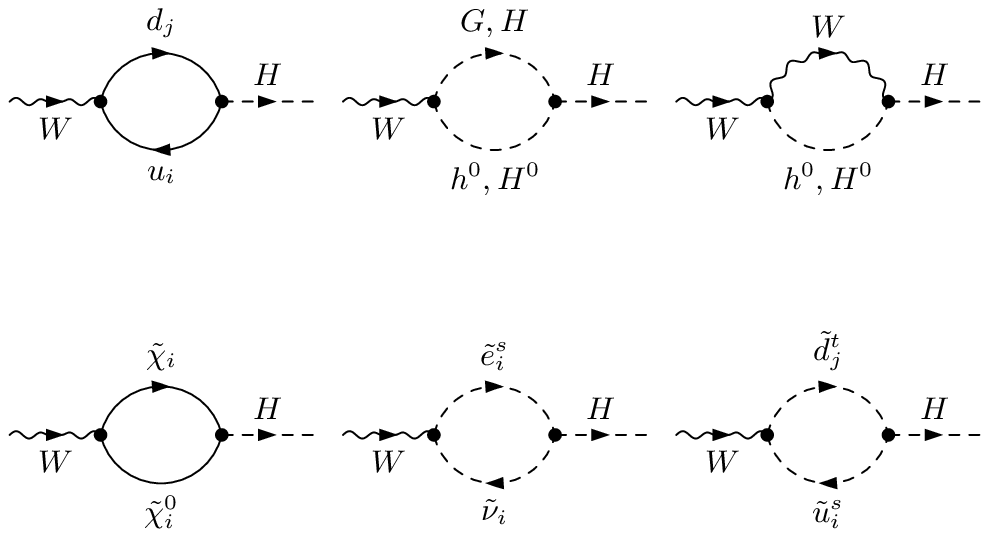}}

(b) $W^+$--$H^-$ self-energy\\[.5cm]
\centering{\includegraphics{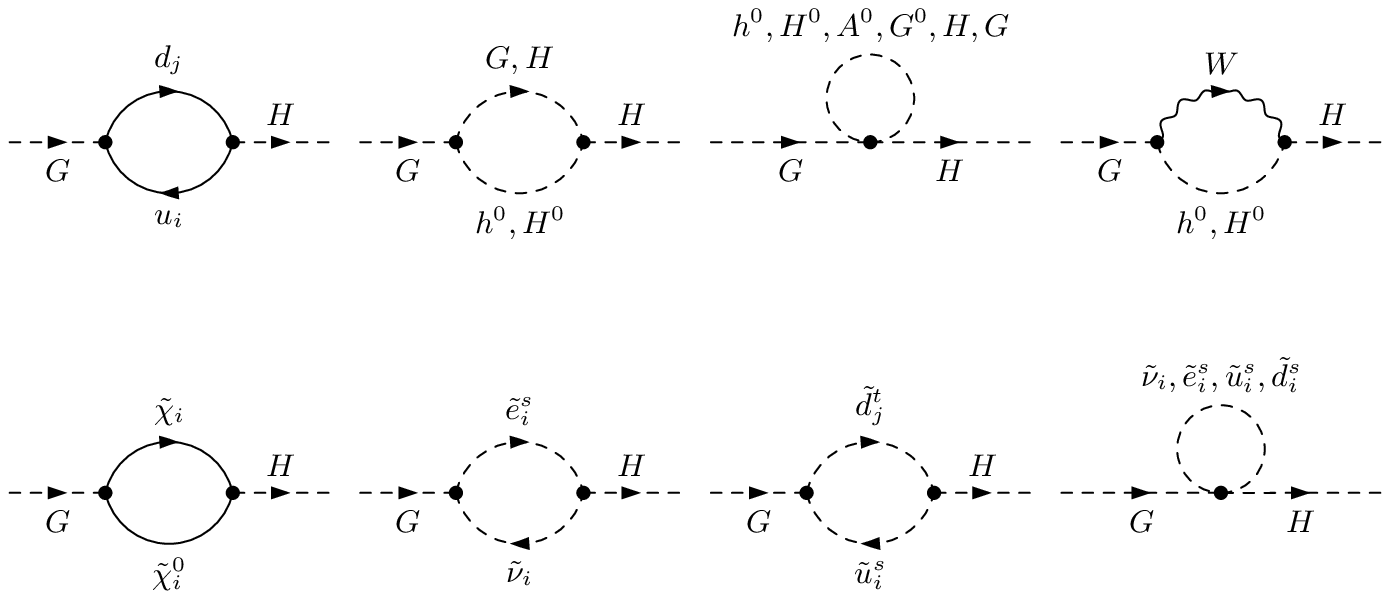}}

(c) $G^+$--$H^-$ self-energy
\caption{\label{fig:eewh-se}
	Feynman graphs for the process $\ee \to W^+ H^-$ with
	self-energy insertions. 
	Each combination of particle labels
	corresponds to a separate Feynman graph.
        }
\end{center}
\end{figure}

\begin{figure}
\begin{center}
\centering{\includegraphics{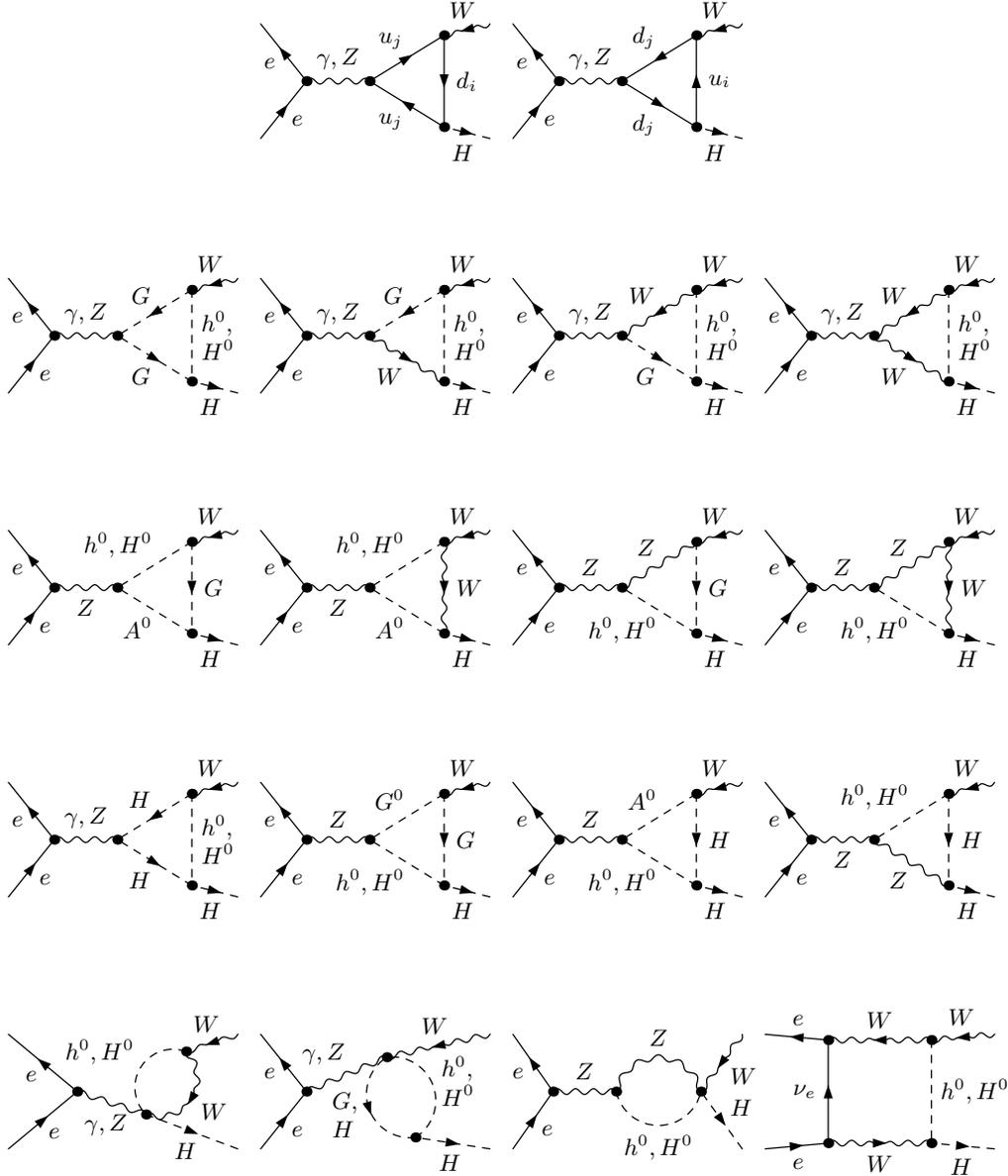}}
\caption{\label{fig:eewh-tribox-2hdm}
	The THDM subset of Feynman graphs of vertex and box type for 
	the process $\ee \to W^+ H^-$.
	Each combination of particle labels
        corresponds to a separate Feynman graph.
        }
\end{center}
\end{figure}

\begin{figure}
\begin{center}
\centering{\includegraphics{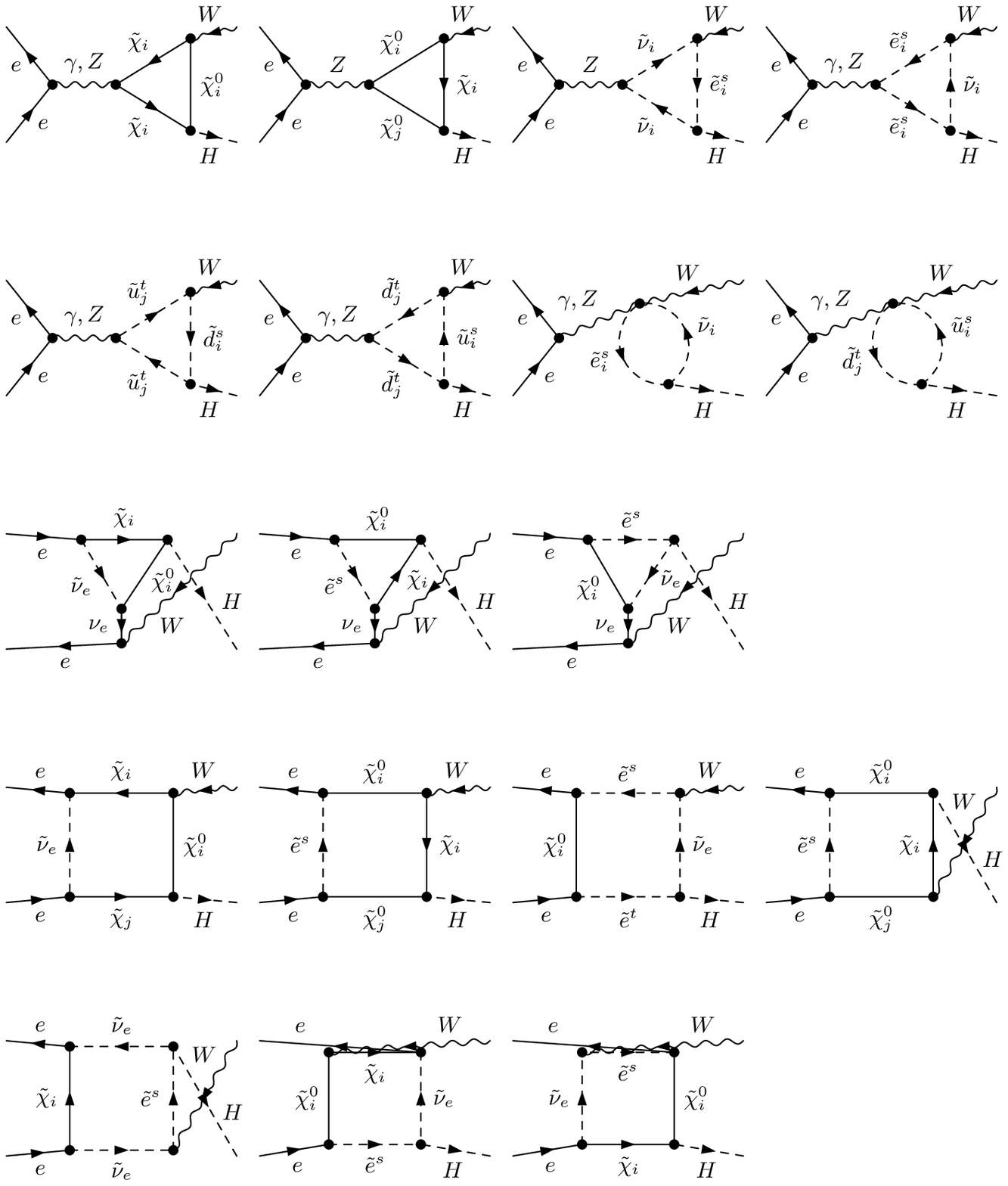}}
\caption{\label{fig:eewh-tribox-susy}
	The superpartner-loop subset of of vertex and box type for
        the process $\ee\to W^+ H^-$.
	Each combination of particle labels
        corresponds to a separate Feynman graph.
        }
\end{center}
\end{figure}

In our calculation we take into account all one-loop contributions to
the amplitude which do not vanish in the limit $m_e = 0$.  For this
reason, Feynman graphs with insertion of a $s$-channel $Z$--$A$ mixing
self-energy, or a $t$-channel neutrino self-energy, or radiative
corrections to $\ee\{h^0,H^0,A^0\}$ Yukawa couplings or to the
$e^\pm\nu_e H^\mp$ Yukawa coupling need not be considered.  There remain
Feynman graphs with insertions of a $W^-$--$H^-$ or $G^-$--$H^-$ mixing
self-energy at the external leg of the outgoing charged Higgs boson (see
Fig.~\ref{fig:eewh-se}), graphs containing the loop-induced $\gamma W^+
H^-$ and $Z W^+ H^-$ vertices\footnote{%
	Note that there is no tree-level $Z W^\pm H^\mp$ coupling in 
	the MSSM or THDM \cite{grifols-mendez}.}
and box-type graphs (see Fig.~\ref{fig:eewh-tribox-2hdm} and
\ref{fig:eewh-tribox-susy}).  The amplitude divides into Feynman graphs
which contain either solely THDM particles (Fig.~\ref{fig:eewh-se}b,c
upper lines and Fig.~\ref{fig:eewh-tribox-2hdm}), or solely
superpartners (Fig.~\ref{fig:eewh-se}b,c lower lines and
Fig.~\ref{fig:eewh-tribox-susy}) in the loop.


\subsection{Calculation}

Although the tree-level contribution vanishes in the the limit of
vanishing electron mass, which we consider, the need for renormalization
arises at one-loop.  There are divergent vector--scalar and
scalar--scalar mixing propagators (Fig.~\ref{fig:eewh-se}) and
vertex-type graphs (Fig.~\ref{fig:eewh-tribox-2hdm} and
\ref{fig:eewh-tribox-susy})\footnote{%
	In principle, also divergent tadpole insertions appear
	which we do not display, because we will choose a 
	renormalization condition such that those contributions vanish 
	(see Eqs.~(\ref{eq:tadpole-cond1}) and 
	(\ref{eq:tadpole-cond2})).}.
We use the on-shell renormalization scheme of
Ref.~\cite{onshell-SUSY-Higgs}, the application of which to
$\wpm$--$\hmp$ self-energies is discussed in detail in
Ref.~\cite{CGGJS}.  
In the MSSM
no new types of divergent vertex functions occur.  Merely, additional
loop contributions with virtual superpartners add to the existing types
of vertex functions.  Therefore, we can take over the renormalization
conditions from Ref.~\cite{arhrib} directly, which are briefly
summarized in the following.

In order to generate the counter-terms, which are needed to renormalize
the amplitude under study, it is sufficient to introduce field
renormalization constants for the MSSM Higgs doublets $H_1$, $H_2$ and
renormalization constants for their vacuum expectation values $v_1$,
$v_2$:
\begin{equation}
  H_i\to \sqrt{Z_{H_i}} H_i\,, 
  \quad
  v_i\to \sqrt{Z_{H_i}} (v_i - \delta v_i)\,,\quad i = 1, 2\,.
\end{equation}
Expanding the renormalization constants in the MSSM Lagrangian
to one-loop order, $Z_{H_i} = 1 + \delta Z_{H_i}^{(1)}$ and
$\delta v_i = \delta v_i^{(1)}$, generates the counter-term interactions
needed and the corresponding Feynman rules:
\begin{align}
\label{eq:ct-wh}
\Gamma_{\text{CT}}[H^\mp W^\pm(k^\mu)] &= 
        \ri \frac{k^\mu}{m_W} m_W^2\, \delta Z_{HW}\,, \\
\Gamma_{\text{CT}}[\gamma_\mu W^\pm_\nu H^\mp] & = 
        -\ri e m_W g_{\mu\nu}\,\delta Z_{HW}\,, \\
\label{eq:ct-zwh}
\Gamma_{\text{CT}}[Z_\mu W^\pm_\nu H^\mp] &=
          \ri e m_W \frac{s_w}{c_w} g_{\mu\nu}\, \delta Z_{HW}
\end{align}
with
\begin{equation}
  \delta Z_{HW} = \sin\beta \cos\beta 
     \biggl( \frac{\delta v_1}{v_1} - \frac{\delta v_2}{v_2} +
     \frac{\delta Z_{H_2} - \delta Z_{H_1}}{2}\biggr)\,,
\end{equation}
where $Z, W^\pm$ und $\gamma$ denote the electroweak gauge bosons
and the photon, and $k^\mu$ the momentum of the $W^\pm$ boson, chosen
as incoming.  In the on-shell scheme the following renormalization 
conditions are posed:
\begin{itemize}
\item
Renormalized tadpole graphs vanish, \ie
\begin{align}
  \label{eq:tadpole-cond1}
  {\hat t}_{h^0} &= t_{h^0} + \delta t_{h^0} = 0\,, \\
  \label{eq:tadpole-cond2}
  {\hat t}_{H^0} &= t_{H^0} + \delta t_{H^0} = 0\,.
\end{align}
This guarantees that the parameters $v_1$, $v_2$ in the renormalized
Lagrangian describe the minimum of the Higgs potential at one-loop 
order.

\item
Real charged Higgs bosons $H^\pm$ do not mix with longitudinally 
polarized $W^\pm$ bosons, \ie the real part of the renormalized 
$H^\pm$--$W^\mp$ mixing self-energy\footnote{
	The renormalized $H^\pm$--$W^\mp$ mixing self-energy is defined
	as the coefficient of $-\ri\frac{k^\mu}{m_W}$ of the amputated
	renormalized $H^\pm$--$W^\mp$ propagator.},
\begin{equation}
  \hat\Sigma_{HW}(k^2) = \Sigma_{HW}(k^2) - m_W^2 \delta Z_{HW}\,,
\end{equation}
vanishes if the momentum $k$ of $H^\pm$ is on mass-shell:
\begin{equation}
  \label{eq:HW-cond}
  \Re\hat\Sigma_{HW}(k^2) \big|_{k^2 = \mhpm^2} = 0\,.
\end{equation}
\end{itemize}
The condition (\ref{eq:HW-cond}) fixes the renormalization constant 
$\delta Z_{HW}$:
\begin{equation}
  \delta Z_{HW} = \frac 1{m_W^2} \Re\Sigma_{HW}(\mhpm^2)\,.
\end{equation}
Because of the BRS symmetry of the renormalized Lagrangian the 
renormalization of the divergent $H^\pm$--$G^\mp$ mixing self-energy
is connected with the $H^\pm$--$W^\mp$ mixing self-energy
through a Slavnov--Taylor identity \cite{CGGJS}:
\begin{equation}
  k^2 \hat\Sigma_{HW} (k^2) - m_W^2 \hat\Sigma_{HG}(k^2) = 0 \;.
\end{equation}
As a consequence, the real part of the renormalized $H^\pm$--$G^\mp$ 
mixing self-energy,
\begin{equation}
  \hat\Sigma_{HG}(k^2) = \Sigma_{HG}(k^2) - k^2 \delta Z_{HG}\,,
\end{equation}
also vanishes for $k^2 = \mhpm^2$:
\begin{equation}
  \Re\hat\Sigma_{HG}(k^2) \big|_{k^2 = \mhpm^2} = 0\,.
\end{equation}
The Feynman rule for the corresponding counter-term interaction 
thus reads:
\begin{equation}
  \Gamma_{\text{CT}}[H^\mp G^\pm (k) ] = \ri k^2 \delta Z_{HG}
\end{equation}
with $\delta Z_{HG} = -\Re\Sigma_{HG}(\mhpm^2)/\mhpm^2$.

The calculation of the amplitude has been performed in 't~Hooft--Feynman
gauge using Constrained Differential Renormalization \cite{CDR} with the
help of the computer programs FeynArts and FormCalc \cite{FAFC}.  To
that end, the counter-term definitions and Feynman rules of counter-term
interactions which were needed were added to the MSSM model file for
FeynArts.

\begin{figure}
\begin{center}
\centering{\includegraphics{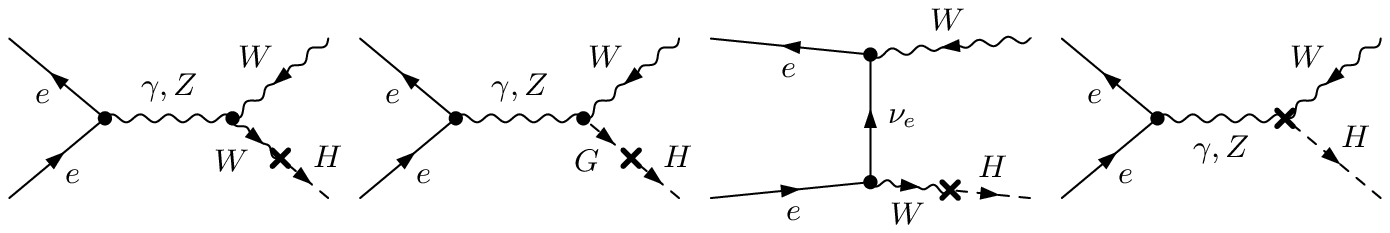}}
\caption{\label{fig:eewh-counter}
        Counter-term diagrams for $\ee\to W^+ H^-$.
        }
\end{center}
\end{figure}

\section{Results}
\label{sect:results}

Our presentation of results is meant to be exemplary.  Given the
possible time-line for the realization of a high-energy 
$e^+ e^-$ collider like the International Linear Collier (ILC) 
and given that it is not yet known what the LHC
will reveal about this new energy frontier, there is no point in
attempting a comprehensive presentation.  The predictions for $\WH$
production in the framework of the general THDM are well known
\cite{zhu-eewh,shinya-eewh,arhrib}.  Our concern is to demonstrate
possible distinctive differences between the MSSM and the THDM.  To that
end, we compare the prediction of our complete one-loop calculation in
the MSSM with the THDM prediction with the Higgs-sector parameters
restricted to MSSM values (sTHDM).  The latter model corresponds to the
MSSM with decoupled superpartners.  Thus, one achieves insight in the
contribution of superpartner loop graphs to the cross section.

\subsection{MSSM parameter scenarios}

In order to demonstrate possible differences between the sTHDM and the
full MSSM, we are interested in scenarios which potentially show large
effects from virtual superpartners.  Hence, the MSSM parameter scenarios
for which we present numerical results have a rather low squark-mass
scale.  In the MSSM, large mixing in the stop sector is known to induce
large radiative corrections in the Higgs sector.  Therefore, we
investigate two types of scenarios, one with large mixing in the stop
sector and the other one with no mixing.

In order to specify the relevant MSSM parameters,
let us recall the stop mass matrix.
For real values of the supersymmetric Higgsino mass parameter $\mu$ 
and the soft-breaking parameter $A_t$
the stop mass matrix reads in the L--R basis
\begin{equation}
  M_{\Stop}^2 = \begin{pmatrix}
    \MSQ^2 + m_t^2 + m_Z^2 (\frac 12 - Q_t s_w^2)\cos 2\beta &
    m_t X_t \\
    m_t X_t &
    \MSU^2 + m_t^2 + m_Z^2 Q_t s_w^2\cos 2\beta
  \end{pmatrix}
\end{equation}
where $X_t = A_t - \mu\cot\beta$, $Q_t = 2/3$, and $s_w = \sin\theta_w$. 
$\MSQ$ and $\MSU$ are the soft-breaking masses for squark isospin-doublets 
and -singlets respectively.
The two MSSM parameter scenarios are specified as follows:

\begin{description}
\item{{\em Large-stop-mixing scenario}:}
We set all sfermion mass parameters ($\MSQ, \MSU, \MSD, \MSL, \MSE$) equal
to one common sfermion mass scale $\MSf = 350\,\gev$,
the stop mixing parameter $X_t  = -806\,\gev$,
$\mu  = 300\,\gev$,
the gaugino mass parameters $M_2 = 1000\,\gev$ and 
$M_1 = (5 s_w^2/3 c_w^2)M_2 = 476.26\,\gev$ (GUT relation).

The main features of this scenario are:
\begin{itemize}
\item The mass of the light stop is of order $100\,\gev$ and there
is a large mass splitting between the two stop mass eigenstates.

\item The stop mixing angle is approximately $45^\circ$.  Thus,
maximal mixing occurs, \ie all entries in the stop mixing matrix have
approximately the same magnitude.

\item The mass of the lightest MSSM Higgs boson $m_h$ is almost maximal
with respect to $X_t$ (because $|X_t|\approx 2\MSf$ \cite{HHW}), \eg
for $\mhpm = 350\,\gev$ and $\tb = 30$ we have $m_h = 120.06\,\gev$.

\item This scenario allows for the production of stop pairs at the
energy scales considered in the following sections and the decay of the
charged Higgs into stop and sbottom will be kinematically allowed for
$\mhpm\gtrsim 450\,\gev$.  
\end{itemize}
\item{{\em No-stop-mixing scenario}:}
Here, we choose the same values of the MSSM
parameters as above, except for $X_t$ which is set to zero.

The main features of this scenario are:
\begin{itemize}
\item Both stops have masses around $400\,\gev$
and all other sfermions have masses around $350\,\gev$.

\item The stop mixing angle is zero.  Thus,
no mixing occurs, \ie the L--R basis for stops 
coincides with the mass basis. 

\item 
The  mass of the lightest MSSM Higgs boson is rather low, 
\eg $m_h = 103.02\,\gev$ for $\mhpm = 350\,\gev$ and $\tb = 30$.
\end{itemize}
\end{description}
For the values of $\mhpm$ and $\tb$ chosen for our discussion, 
both scenarios respect all
superpartner mass bounds from direct search results and fulfil indirect
constraints on the parameters coming from requiring vacuum stability and
from experimental bounds on the electroweak precision observables.
Also the lower bound on the lightest neutral MSSM Higgs boson is respected
by both scenarios, except for the case $\tb=3$ in the no-mixing scenario.
Yet, for the sake of comparison, we also show this case among our numerical results.

\subsection{Cross-section for the parameter scenarios}

There are two major motivations for studying the MSSM prediction for
$\ee\to\WH$ cross sections at a future ILC.
Firstly, it is important to know the expected event rates
at the ILC either to confirm the discovery of charged Higgs bosons at
the LHC or to assess the $H^\pm$ discovery potential of the ILC beyond
the pair-production limit of $\sqrt s/2$.  
Secondly, the observation of
this process provides the
opportunity to glean some information
about the underlying model 
because of its potential sensitivity to virtual superpartners.

We present results for two collider energies, $\sqrt s = 500\,\gev$ and
$1000\,\gev$. For each energy we show results for the the large-stop-mixing 
and no-stop-mixing scenario 
for different $\tb$-values in and around the ``wedge region''
and for charged Higgs bosons masses below, above and right at the pair
production limit of 250 and 500 GeV, respectively.

\subsubsection{Large-stop-mixing scenario}

In Figs.~\ref{fig:mhpm-eehw}a and \ref{fig:tb-eehw}a the $\ee\to\WH$
cross sections for a collider energy of $500\,\gev$ are displayed for the
full MSSM and the corresponding sTHDM.
Fig.~\ref{fig:mhpm-eehw}a shows the $\mhpm$-dependence for $\tb = 3$,
$15$, and $30$.  The spike in all predictions for $\mhpm\equiv\sqrt{\bar
p^2}\approx 180\,\gev\approx m_t+m_b$ is due to a threshold effect in the
vertex graphs with virtual top- and bottom-quarks (see
Fig.~\ref{fig:eewh-tribox-2hdm}).  Similar spikes appear in the full
MSSM prediction in places where $\mhpm$ equals $m_{\Stop_i} +
m_{\Sbot_j} (i,j = 1,2)$ and are due to threshold effects in the Feynman
graphs with virtual stops and sbottoms, \eg for $\tb = 30$ where
$\mhpm\approx 390\,\gev\approx m_{\Stop_1}+m_{\Sbot_1}$.
Fig.~\ref{fig:tb-eehw}a shows the $\tb$-dependence for $\mhpm =
190\,\gev$, $250\,\gev$, and $350\,\gev$.  The MSSM cross section is generally
about two orders of magnitude larger than the sTHDM in the large-stop-mixing case for
large values of $\tb$.

For $\sqrt s = 1000\,\gev$ Figs.~\ref{fig:mhpm-eehw}b and
\ref{fig:tb-eehw}b reveal the same general behaviour.  Yet, due to the
higher collider energy, more stop--sbottom thresholds are within the
accessible mass range for the charged Higgs boson.  

The enhancement with
respect to the sTHDM case is due to the threshold effects combined with
enhanced couplings of the charged Higgs boson to third-generation
squarks.
In the large-stop-mixing scenario, the Feynman graphs with virtual squarks of the
third generation comprise the dominant contribution to the cross section
if $\tb$ is not too small.  In this scenario each entry of the rotation
matrix which transforms the stop fields into the mass eigenbasis is of
similar magnitude.  Thus, all potentially large terms in the couplings
of third-generation squarks to Higgs bosons\footnote{%
	A full list of the relevant couplings in our notation can be 
	found in \cite{gghphm}}
are weighted by a factor of order 1, especially the term proportional
to $m_t\mu$.

\subsubsection{No-stop-mixing scenario}

\begin{figure}[hbt]
\centerline{\includegraphics{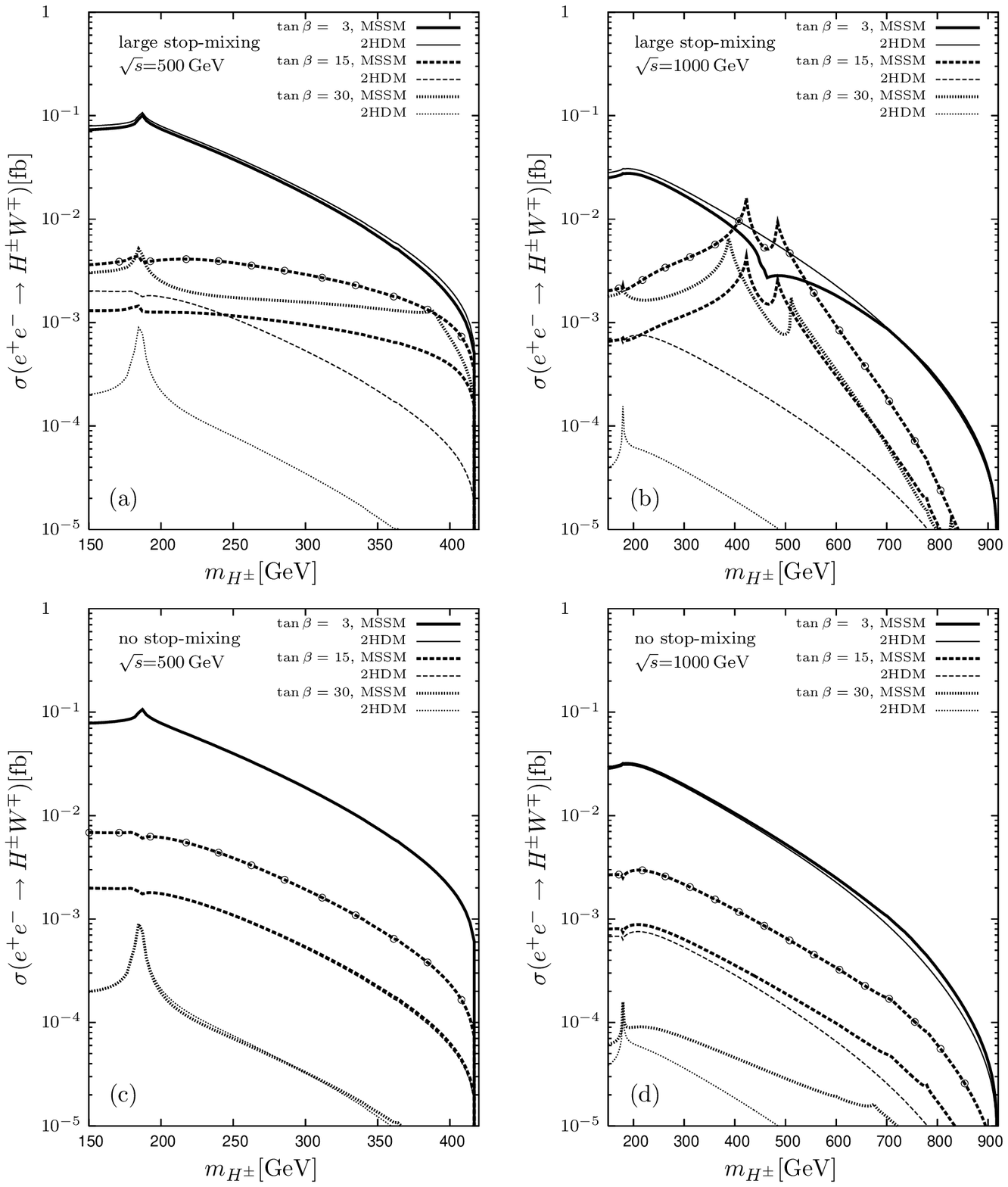}}
    \caption{
	\label{fig:mhpm-eehw}
	Cross section $\sigma(\ee \to H^\pm W^\mp)$ versus 
	$\mhpm$ for different values of $\tb$.
	MSSM predictions 
	of the large-stop-mixing and no-stop-mixing scenarios
	for the unpolarized cross section
	are displayed (thick lines) 
	along with the corresponding sTHDM scenario (thin lines)
	for a collider energy of $\sqrts = 500\,\gev$
	and $\sqrts = 1000\,\gev$. The cross section for optimal
	polarization, $P(e^-)=-1$ and $P(e^+)=+1$, is shown
        for the case $\tb=15$ (thick lines with circles).
        }
\end{figure}

\begin{figure}[hbt]
\centerline{\includegraphics{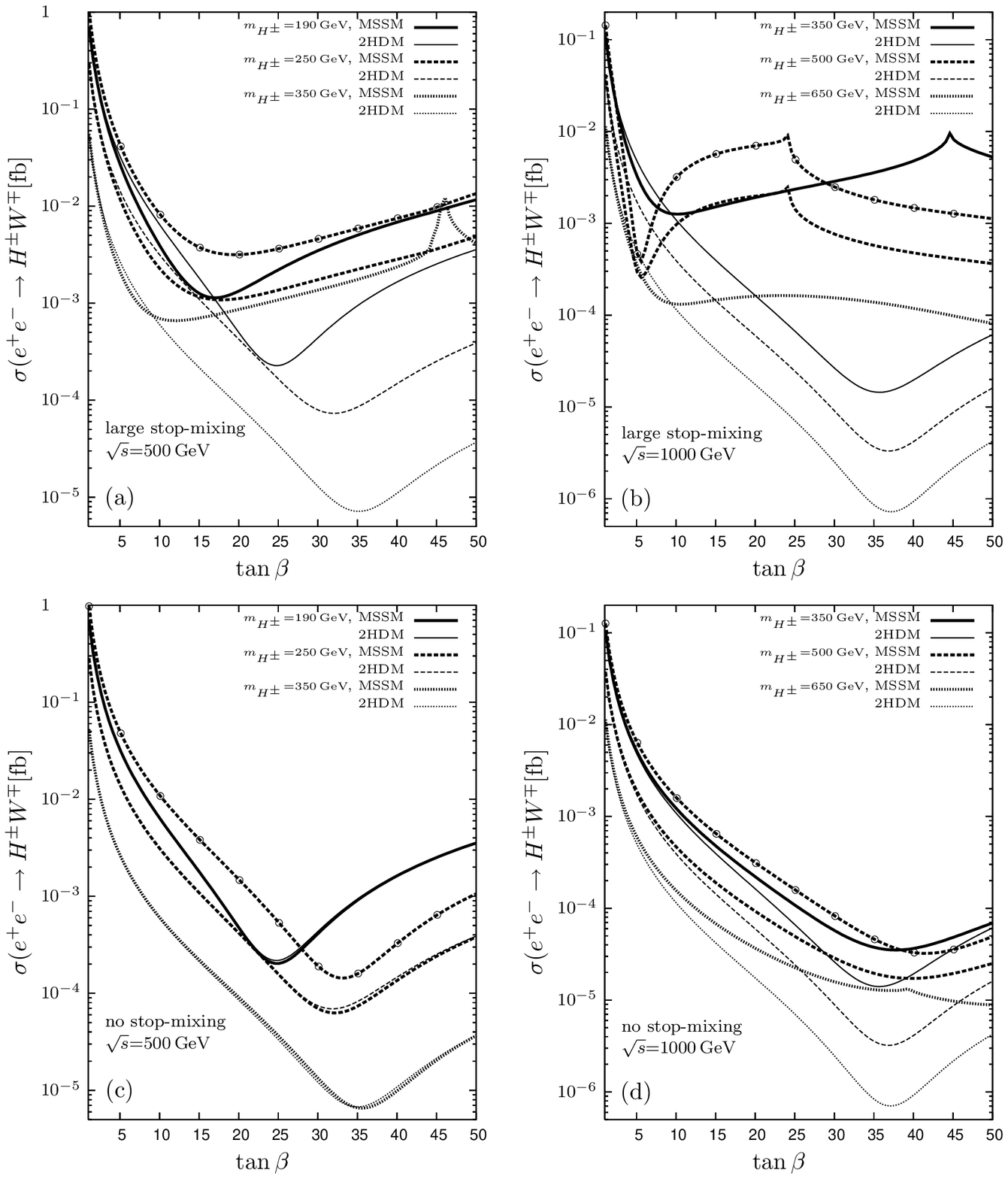}}
    \caption{
	\label{fig:tb-eehw}
	Cross section $\sigma(\ee \to H^\pm W^\mp)$ versus $\tb$ 
	for different values of $\mhpm$.
	MSSM predictions 
	of the large-stop-mixing and no-stop-mixing scenarios
	for the unpolarized cross section
	are displayed (thick lines) 
	along with the corresponding sTHDM scenario (thin lines)
	for a collider energy of $\sqrts = 500\,\gev$
	and $\sqrts = 1000\,\gev$. The cross section for optimal
	polarization, $P(e^-)=-1$ and $P(e^+)=+1$, is shown in each panel
        for the case $\mhpm=\sqrts/2$ (thick lines with circles).
        }
\end{figure}

Figs.~\ref{fig:mhpm-eehw}c and \ref{fig:tb-eehw}c show the predicted
cross section for a $500\,\gev$ collider in this scenario.  Superpartner
effects are small in this case and the MSSM as well as the sTHDM predict
almost the same cross section.  Without mixing in the stop sector the
two factors which caused the large enhancement in the large-mixing case
are absent.
Firstly, there are no enhancement effects by thresholds in 
stop--sbottom loops.
Secondly, the mixing angle in the stop sector is exactly zero which
leads to the absence of terms proportional to $m_t\mu$ in the
$\Stop_1\Sbot_i H^\pm$-couplings.

For $\sqrt s = 1000\,\gev$ the situation changes (see
Figs.~\ref{fig:mhpm-eehw}d and \ref{fig:tb-eehw}d).  The maximally
accessible charged-Higgs-boson mass is higher than all thresholds in
stop--sbottom loop graphs ($730\,\gev < m_{\Stop_i} + m_{\Sbot_j} <
760\,\gev$).  Again, there are threshold enhancements of the MSSM
prediction with respect to the sTHDM of up to one order of magnitude for
large $\tb$.

\subsubsection{Effect of polarized electrons and positrons}

The cross section of $\ee\to\WH$ depends strongly on the 
polarization of the incoming electrons and positrons. 
Specifically, the cross section varies roughly by
one order of magnitude between the optimal situation 
where all electrons have helicity $-1$
and all positrons have helicity $+1$, 
i.e. $P(e^-)=-1$ and $P(e^+)=+1$, and the opposite situation
where $P(e^-)=+1$ and $P(e^+)=-1$.

In order to illustrate the effect of polarized beams at a future LC,
we include in Figs.~\ref{fig:mhpm-eehw}a to \ref{fig:mhpm-eehw}d 
the cross section
prediction for the optimal polarization for $\tb=15$. Likewise,
we include in Figs.~\ref{fig:tb-eehw}a to \ref{fig:tb-eehw}d the cross section
prediction for the optimal polarization for $\mhpm=\sqrt s/2$.
Optimal polarization leads to an 
increase in cross section varying between a factor of 
2 up to 3.8
depending on  $\mhpm$ and $\tb$ for both collider energies.

\subsubsection{Decoupling of superpartners}

We set all soft-breaking parameters which appear in our expression for
the MSSM cross section equal to a single supersymmetry breaking scale
$\widetilde M$,
\begin{equation}
  \widetilde M = \MSf = A_b = A_t = M_1 = M_2 = \mu \,,
\end{equation}
and study the behaviour of the MSSM cross section with rising
$\widetilde M$ compared to the sTHDM prediction.  This will allow us to
demonstrate the decoupling of the virtual superpartner contributions to
the cross section.

\begin{figure}[hbt]
\centerline{\includegraphics{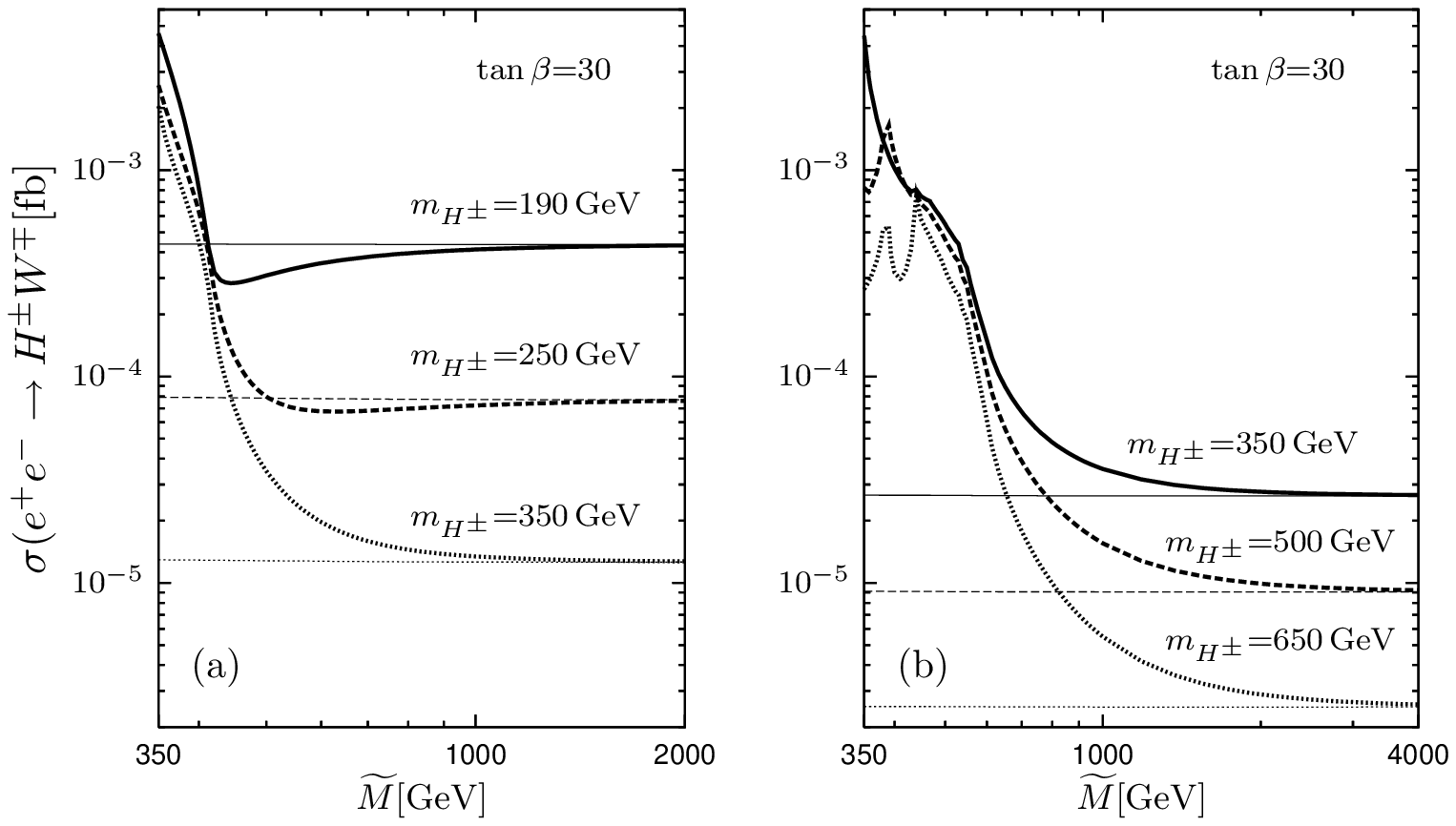}}
    \caption{
	\label{fig:decoupling-plot}
	Cross section for the process 
        $\ee \to H^\pm W^\mp$ as a function
	of a common superpartner mass scale $\widetilde M$
	for a (a) $500\,\gev$ and (b) $1\,\tev$ $e^+ e^-$
 	collider.
        }
\end{figure}

Fig.~\ref{fig:decoupling-plot}a shows results for a $500\,\gev$ collider
and $\mhpm = 190$, $250$, and $350\,\gev$ with $\widetilde M$ varied
between $350$ and $2000\,\gev$.  The MSSM results drop significantly once
all superpartners get heavy enough for threshold effects to disappear. 
A similar behaviour is shown in Fig.~\ref{fig:decoupling-plot}b for a
$1000\,\gev$ collider and $\mhpm = 350$, $500$, and $650\,\gev$ where
$\widetilde M$ is varied up to $4\tev$.  Notably, even at a
soft-breaking scale of $1\tev$, the MSSM surpasses the sTHDM prediction 
by a factor of 1.4, 1.7, and 2.2 for these three mass values, 
respectively.

\subsection{Parameter scan}

We demonstrated the possibility of strong enhancement of the MSSM
cross section prediction by virtual superpartner effects for the
large-stop-mixing scenario.  
This might predict $\WH$ production to be observable in
such MSSM scenarios unlike in the corresponding sTHDM scenario.  We
investigate this possible behaviour more systematically by scanning over
the relevant MSSM parameter space.

The detectors at the LHC will be capable of 
finding an MSSM charged Higgs boson only in
certain regions in the $\mA$--$\tb$ plane.  
Likewise, the ILC can detect charged Higgs bosons easily only 
if pair production
is kinematically possible.  Thus, as far as $\WH$ production at the ILC
is concerned, there are two points of departure: either we will know to
some extent what the mass of the charged Higgs boson is or we will not.
In the first case it makes sense to combine this knowledge with the
knowledge of the cross section prediction in the MSSM and adjust the
collider energy to the maximum of the expected cross section as a
function of $\sqrt s$.
In the second case it is difficult to say what the best strategy of
choosing $\sqrt s$ will be for searching for the charged Higgs boson. 
A reasonable assumption is however that a significant fraction of the
data during ILC operation will be collected at the highest collider
energy.

With these two possibilities in mind, we devised two parameter scans. 
The first scan assumes that the charged Higgs mass is known and that it
makes sense to adjust the collider energy to maximize the cross section. 
Furthermore, it assumes that one has also acquired some knowledge about
the value of $\tb$.  We thus fix $\mhpm$ and $\tb$ in our scan and
include $\sqrt s$ in the list of scanned parameters.
The second scan fixes $\sqrt s$ at the maximal collider energy but
includes $\mhpm$ and $\tb$ in the list of scanned parameters.  For both 
scans we assume an ILC with a maximal collision energy of 1 TeV.

In general, we are interested in regions of parameter space 
where the cross section is large
such that it might still be detectable at the ILC.
In order to increase the number of scanned parameter points in the region 
of large cross section, we use the ''adaptive scanning'' algorithm
described in more detail in Ref. \cite{scan-note}.
Essentially, that means that we are using the
importance sampling 
algorithm VEGAS \cite{vegas} to evaluate the 
integral of the cross section over the relevant
MSSM parameters.
However, we are not interested in the value of the meaningless integral
but store the sampled points which automatically accumulate in the
region of large cross section.
This method 
allows to perform scans of higher-dimensional parameter spaces
with emphasis on certain features of the results
much more efficiently than a grid of points or purely random sampling
can do. 
More precisely, the quantity we are integrating over
in our scans 
is  identical to the cross section only at `allowed' points according to the
exclusion limits defined below, and zero otherwise.  
The enrichment of points thus focuses also on the allowed region.
We use the following exclusion limits:
\begin{equation}
\begin{aligned}
m_{\Stop_i} &\geqslant 95.7\,\gev\,, \qquad &
m_{\Sbot_i} &\geqslant 89\,\gev\,, &
m_{\Sq\neq \Sbot, \Stop}
  &\geqslant 150\,\gev\,, \\
m_{\tilde\chi^0} &\geqslant 46\,\gev\,, &
m_{\tilde\chi^\pm} &\geqslant 94\,\gev\,, &
M_{\tilde\ell} &\geqslant 73\,\gev\,, \\
& &  
|\Delta\rho| &\leqslant 0.0025,
\end{aligned}
\end{equation}
where $\Delta\rho$ is the dominant squark contribution
to the electroweak rho-parameter.
The mass bounds on superpartners 
and on $\Delta\rho$
are according to Ref.~\cite{pdg2004}.

Apart from these exclusion limits, we take into account two more 
major constraints on the MSSM parameter space.

Firstly, we calculate for each parameter point the MSSM predictions for 
$m_{h^0}$ and $\sigma(e^+ e^- \to h^0 Z)\times\BR(h^0\to b\bar b)$
and exclude it if the $m_{h^0}$-dependent LEP-bound on $\sigma\times\BR$
is violated (according to Table 14(b) of Ref.~\cite{LEP-susy-Higgs}).
We use FeynHiggs 2.5.1 \cite{FH} for calculating the $m_{h^0}$ prediction
and allow for a theoretical uncertainty of $3 \,\gev$.

Secondly, we calculate the leading-order MSSM prediction for the 
branching ratio $\BR(B \to X_s \gamma)$ \cite{bsg-mssm} and exclude parameter points
if the prediction falls outside of the range $(3.55 \pm 1.71)\TE{-4}$.
This range is determined by using the experimental central value
\cite{bsg-exp} and 
adding up the experimental $3\sigma$ interval ($\approx \E{-4}$) 
and an estimate of the independent theoretical uncertainty ($0.71\TE{-4}$). 
The latter estimate (20\%) is guided by the 
discussion of theoretical uncertainties for the SM prediction
\cite{misiak-etal}.

\subsubsection{Scan 1: $\mhpm$ and $\tb$ known}

\begin{figure}[hbt]
\centerline{\includegraphics{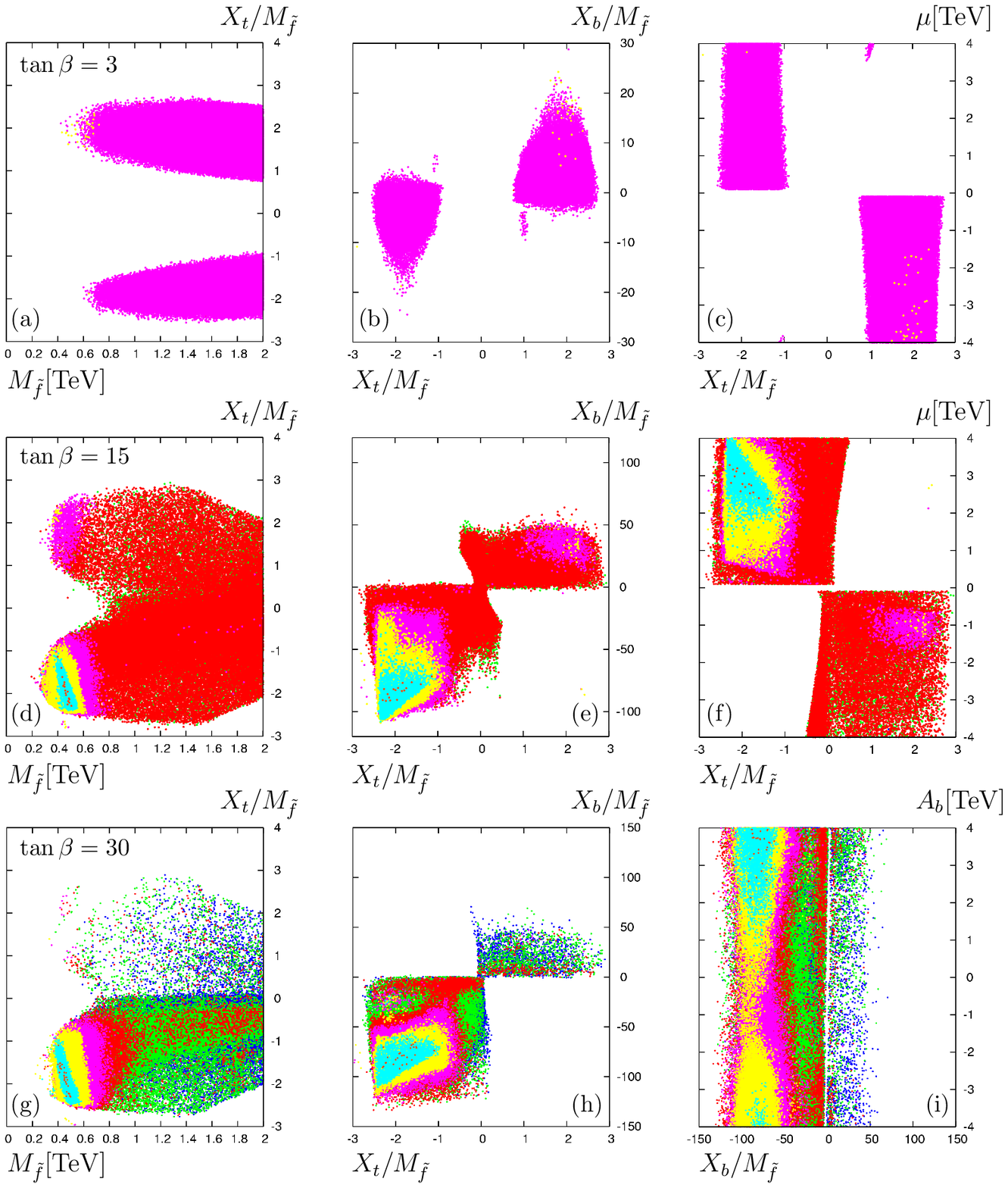}}
    \caption{
	\label{fig:scan1-2d-slices}
	Two dimensional slices through the scanned parameter space of Scan 1
	for $\tb=3, 15$ and $30$.
	The colours refer to bins of cross section values:
	\GNUPlotH{{\bf orange}}: $10\fb \geq \sigma >  1\fb$,
	\GNUPlotE{{\bf light blue}}: $1\fb \geq \sigma > \E{-1}\fb$,
	\GNUPlotF{{\bf yellow}}: $\E{-1}\fb \geq \sigma > \E{-2}\fb$,
	\GNUPlotD{{\bf cyan}}: $\E{-2}\fb \geq \sigma > \E{-3}\fb$,
	\GNUPlotA{{\bf red}}: $\E{-3}\fb \geq \sigma > \E{-4}\fb$,
	\GNUPlotB{{\bf green}}: $\E{-4}\fb \geq \sigma > \E{-5}\fb$,
	\GNUPlotC{{\bf blue}}: $\sigma < \E{-5}\fb$.
	Points with larger values of $\sigma$ are plotted on top of points with smaller 
	values.
        }
\end{figure}

The MSSM input parameters are scanned over the following region:
\begin{equation}
  \begin{aligned}
  \MSf, M_1, M_2 &= 10 \ldots 2000\,\gev\,, \\
  \mu, A_t, A_b &= -4000 \ldots 4000\,\gev\,, \\
  \sqrt s &= 500\ldots 1000\,\gev\,,\\
  \mhpm &= 500\,\gev\,, \\
  \tb &= 3,\ 15,\ 30\,.
  \end{aligned}
\end{equation}
The $\tb$-values respectively lie roughly at the lower, middle, and
upper end of the wedge-region.  Examining the results by studying
projections of the cross section on all one- and two-dimensional
subspaces of the scanned parameter space we find that the gaugino mass
parameters $M_1$ and $M_2$ have negligible influence on the
cross section, at least in all regions where the cross section is above
$\E{-3}\,\fb$.  Because of their importance we also study the
dependence of the cross section on the squark mixing parameters $X_t, X_b$
normalized to the sfermion mass scale $\MSf$, i.e. 
$\hat X_t = (A_t - \mu/\tb)/\MSf$ and $\hat X_b = (A_b - \mu\tb)/\MSf$.  In
Fig.~\ref{fig:scan1-2d-slices} we show the most interesting
two-dimensional projections \cite{scan-web}.

For $\tb=3$ the most noticeable feature, shown in 
Figs.~\ref{fig:scan1-2d-slices}a, \ref{fig:scan1-2d-slices}b and \ref{fig:scan1-2d-slices}c,
is that the highest values for the $H^\pm W^\mp$ production cross section 
lie typically between $\E{-3}\,\fb$ and $\E{-2}\,\fb$ and are reached 
everywhere in the allowed parameter region.
In our scan we find a few cases with higher cross section, clearly visible as the 
scattered yellow dots on a cyan background. 
Yet, we find no scenario with a cross section above 0.1$\,\fb$.
The well-known fact that for small $\tb$
the constraints, especially the LEP Higgs-mass bound, require a fair amount of 
stop mixing (i.e. $|\hat X_t| \gtrsim 1$) to be present, 
is clearly seen in all three figures.

Also owing to the LEP bound, the lowest common squark mass scale $\MSf$ for which 
we find allowed scenarios is about $400\,\gev$ for positive $\hat X_t$ and
$600\,\gev$ for negative $\hat X_t$.
It is for rather low $\MSf$, below about $650\,\gev$, that the few scenarios
with a cross section above $\E{-2}\,\fb$ appear.
It turns out that having $\MSf \lesssim 650\,\gev$ is a 
rather generic necessary requirement for scenarios with cross sections 
above $\E{-2}\,\fb$, as will become more clear in the following.

The projection of the parameter points on the $\hat X_t$--$\mu$ plane 
in Fig.~\ref{fig:scan1-2d-slices}c reveals that 
almost all of the scenarios we find in our scan show
a strict correlation between the sign of $\hat X_t$ and the sign of $\mu$,
the former being positive if the latter is negative and vice versa.

For $\tb=15$ 
(Figs.~\ref{fig:scan1-2d-slices}d, \ref{fig:scan1-2d-slices}e and \ref{fig:scan1-2d-slices}f) 
the results of the scan
change drastically compared to the $\tb=3$ case. 
Firstly, the allowed region in the $\MSf$--$\hat X_t$ plane 
(see Fig.~\ref{fig:scan1-2d-slices}d) is much larger, now also including 
scenarios without stop mixing and scenarios with $\MSf$ below $300\,\gev$.
Secondly, the highest cross section values 
are considerably larger, of the order of $10\,\fb$.
Interestingly, the projection of the parameter points on the $\MSf$--$\hat X_t$ plane
(Fig.~\ref{fig:scan1-2d-slices}d) shows that
the scenarios with a (large) cross section above $1\,\fb$
are confined to a very specific region with 
$\MSf$ between $250\,\gev$ and $600\,\gev$, 
and $\hat X_t$ between $-2.5$ and $-1$ simultaneously.

The projection on the $\hat X_t$--$\hat X_b$ plane (Fig.~\ref{fig:scan1-2d-slices}e)
shows that all large cross section scenarios also have a significantly nonzero and
negative value of $\hat X_b$. 
Quantitatively, it turns out that those scenarios typically have the two mass scales
$m_t \hat X_t$ and $m_b \hat X_b$ of the same order of magnitude.
Furthermore, from the projection on the $\hat X_t$--$\mu$ plane (Fig.~\ref{fig:scan1-2d-slices}f) 
we see that those scenarios also have a positive and large value of $\mu$, 
typically between $2\,\tev$ and $4\,\tev$ (the limit of our scan range). 
Looking into the $\hat X_b$-$A_b$ plane (not depicted) we find large cross section
scenarios for almost any value of $A_b$.

The scan results for $\tb=30$ 
(Figs.~\ref{fig:scan1-2d-slices}g, \ref{fig:scan1-2d-slices}h and \ref{fig:scan1-2d-slices}i)
look very similar to the previous case. 
Again, in order to obtain a cross section above $0.1\,\fb$ it appears that
one needs a low sfermion mass scale $\MSf$ between roughly $250\,\gev$ and $600\,\gev$
and a significant amount of mixing in the stop {\em and} sbottom sector
with both $\hat X_t$ and $\hat X_b$ negative.
One important change compared to the previous case is that all large cross section
scenarios have a rather large value 
of $|A_b| \gtrsim 1\,\tev$ (see Fig.~\ref{fig:scan1-2d-slices}i)
while the values of $\mu$ are constrained to $0.8\,\tev \lesssim \mu \lesssim 2.2\,\tev$
(result not depicted).

\subsubsection{Scan 2: $\mhpm$ and $\tb$ unknown}

The MSSM input parameters are scanned over the following region:
\begin{equation}
\label{scan2}
  \begin{aligned}
  \MSf &= 10\ldots 2000\,\gev\,, \\
  \mu, A_t, A_b &= -4000\ldots 4000\,\gev\,, \\
  \mhpm &= 500\ldots 920\,\gev\,, \\
  \tb &= 1\ldots 50\,,\\
  M_1 = M_2 &= 500\,\gev\,, \\
  \sqrt s & = 1\,\tev\,.
  \end{aligned}
\end{equation}
The less influential gaugino-mass parameters $M_1$, $M_2$ have been
fixed in order to obtain
a more thorough scan over the relevant MSSM parameters.

\begin{figure}[hbt]
\centerline{\includegraphics{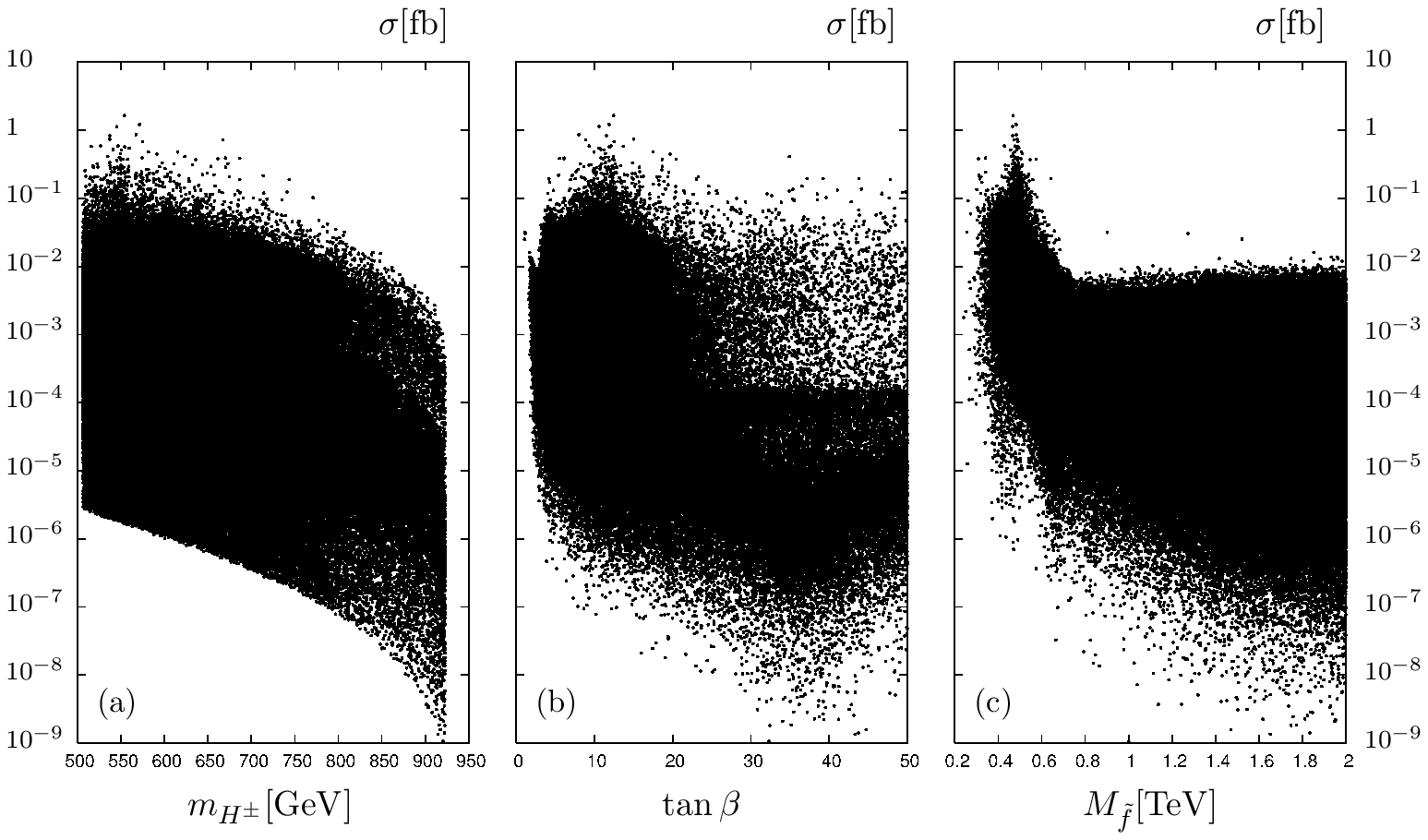}}
    \caption{
	\label{fig:scan2-1d-slices}
	Cross section results of Scan 2 versus (a) $\mhpm$,
	(b) $\tb$ and (c) $\MSf$.
        }
\end{figure}

Fig.~\ref{fig:scan2-1d-slices}a shows
the obtained cross section values projected on the $\mhpm$-axis.
Overall, the values drop with rising $\mhpm$ due to decreasing
phase space.
There are no scenarios with $\sigma > 0.1\,\fb$ for $\mhpm \gtrsim 750\,\gev$
and no scenarios with $\sigma > 0.01\,\fb$ for $\mhpm \gtrsim 875\,\gev$.
The projection on the $\tb$-axis in
Fig.~\ref{fig:scan2-1d-slices}b
shows that there exist allowed parameter scenarios with cross sections
even above $0.1\,\fb$ over the whole scanned range.
Fig.~\ref{fig:scan2-1d-slices}c shows
the very distinct dependence of the cross section 
on the common sfermion mass scale $\MSf$ 
which survives even 
after the inclusion of $\mhpm$ and $\tb$ in the scan.
Evidently, cross sections significantly above $0.1\,\fb$ are only
possible if $275\,\gev\lesssim \MSf \lesssim 700\,\gev$.
A detection of squarks with masses in this range at the LHC 
would raise hopes that $\WH$ production would be
observable at the ILC with an integrated luminosity of 
the order of 1000 events/$\fb$.

\begin{figure}[hbt]
\centerline{\includegraphics{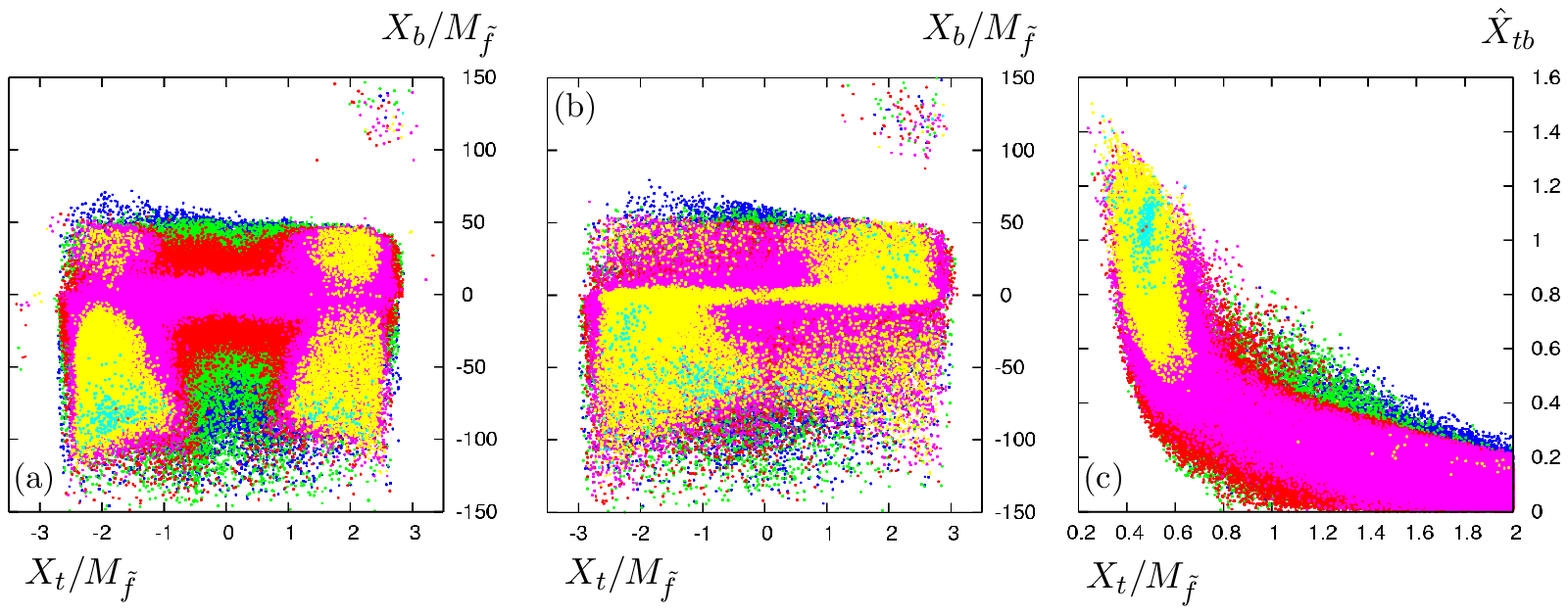}}
    \caption{
	\label{fig:scan2-2d-slices}
	Two dimensional slices through the scanned parameter space of Scan 2.
	The middle panel (b) shows the same slice as panel (a) but 
	for a variant of Scan 2
	which uses a simple implementation of the LEP constraint (see text).
	The quantity $\hat X_{tb}$ in panel (c) is defined in Eq.~(\ref{defxtb}).
	The colours refer to bins of cross section values
	in the same way as in Fig.~\ref{fig:scan1-2d-slices}.
	Points with larger values of $\sigma$ are plotted on top of points with smaller 
	values.
        }
\end{figure}

As in Scan 1 it turns out that the mixing parameters
in the third generation squark sector are important
for the understanding of the large cross section regions.
Fig.~\ref{fig:scan2-2d-slices}a shows that the largest cross section 
values lie roughly on two broad stripes with $1 < |\hat X_t| < 2.5$
in the $\hat X_t$--$\hat X_b$ plane
leaving the central part of the plane with values below $0.01\,\fb$.
We find cross section values above $0.1\,\fb$ only for
scenarios which also have a considerable amount of sbottom mixing,
typically $\hat X_b \lesssim -50$.

At this point, it is instructive to compare our scan with a variant version
where the LEP Higgs constraint on the MSSM parameters 
is implemented in its most simple way, 
which is often done in the literature. 
This simple implementation only makes use of the overall, scenario independent,
bound on $\mh$ (e.g. $89.8\,\gev$ \cite{pdg2004}), 
discarding all scenarios which predict $\mh$ to lie below it.
Obviously, this prescription is overly conservative, keeping e.g. 
all SM-like scenarios with masses above $89.8\,\gev$ although in this case
a bound close to the SM limit of about $114.4\,\gev$ \cite{LEP-SM-Higgs-bound} applies.

Fig. \ref{fig:scan2-2d-slices}b shows the results for Scan 2 
using the simple implementation of the LEP Higgs constraint.
The result looks quite different to the detailed implementation of the constraint
in Fig.~\ref{fig:scan2-2d-slices}a.
The largest cross section 
values lie roughly scattered around a 
half-ring in the $\hat X_t$--$\hat X_b$ plane.
In this less restricted version of Scan 2
a certain amount of squark mixing, 
either in the stop or sbottom sector is needed in order to obtain
cross sections above $0.1\,\fb$.

Motivated by this observation we study the dimensionless quantity
\begin{equation}
\label{defxtb}
\hat X_{tb}  := \sqrt{\left(\frac{m_t X_t}{\MSf^2}\right)^2
	+\left(\frac{m_b X_b}{\MSf^2}\right)^2}
\end{equation}
as a discriminating variable for our original Scan 2. 
This quantity is just the off-diagonal entries 
in the stop and sbottom mass matrices, $m_t X_t$ and $m_b X_b$ respectively,
added in quadrature and normalized to the common sfermion mass scale $\MSf$.
Fig.~\ref{fig:scan2-2d-slices}c
shows the projection in the $\MSf$--$\hat X_{tb}$ plane.
Evidently, cross sections above $0.1\,\fb$ 
require $\hat X_{tb} \gtrsim 0.8$ and 
centre roughly around $\hat X_{tb}\approx 1$.

\section{Conclusions}
\label{sect:conclusion}

The production of a charged Higgs boson $H^\pm$ in association with a
an electroweak boson $W^\mp$ in $\ee$ collisions is a loop-induced process.
At the ILC, this process is particularly important for the
charged-Higgs-boson detection if pair production is kinematically
forbidden. Furthermore, one would gain valuable
information about the underlying model by observing this process at the ILC.
We calculated the cross section for this process at one-loop order in
the framework of the MSSM and THDM and investigated the predictions of
both models at an ILC with $500\,\gev$ and $1000\,\gev$ centre-of-mass
energy.
The MSSM scenarios with large stop mixing and low sfermion mass scale,
for which we showed an example, can give rise to a cross section 
which differs from a THDM with identical Higgs 
sector parameters by two orders of magnitude.
Using polarized electron and positron beams can increase the cross section
by a factor of 2 to 4 depending on $\mhpm$ and $\tb$.
We performed a MSSM parameter scan for regions of large cross section
assuming a $1\,\tev$ collider and a charged Higgs boson too heavy to be 
pair-produced at such a machine. 
We find scenarios with a cross section above $0.1\,\fb$ for $\mhpm$ 
up to about $750\,\gev$ in the whole scanned $\tb$-range ($1\ldots 50$).
These scenarios require a sfermion mass scale between $200\,\gev$
and $600\,\gev$ and a certain amount of mixing in the stop 
and sbottom sector. 
The FORTRAN program 'eeWH'  for the calculation of the MSSM and THDM
cross sections including the option to perform parameter scans
can be obtained from one of the authors
\footnote{Please use the e-mail address oliver.brein@durham.ac.uk}.

\section*{Acknowledgements}

We thank Wolfgang Hollik for useful comments. 
Furthermore, we thank
Rachid Benbrik and Sven Heinemeyer for suggesting to improve
on the MSSM parameter restrictions we take into account.

\begin{flushleft}

\end{flushleft}

\end{document}